\documentclass[12pt]{article}
\usepackage{epsfig}
\usepackage{latexsym}
\begin{document}

\title{A possibility to describe models of nonabelian massive gauge fields in the framework of renormalizable theory}
\author{ A.A.Slavnov \thanks{E-mail:$~~$ slavnov@mi.ras.ru}\\
V.A.Steklov Mathematical Institute\\ Gubkina street 8, GSP-1,119991}

\maketitle

\begin{abstract}
Renormalizable theory of massive nonabelian gauge fields, which does not require the existence of observable scalar fields is proposed.
\end{abstract}

\section{ Introduction}

Although the discovery at CERN of the particle which probably may play the role of the scalar meson, predicted in the papers (\cite{BE}, \cite{Hi})and is the foundation of electroweak sector
of the Standard Model (\cite{W},\cite{S}, \cite{G}) partially resolved the question about the validity of the Standard Model, creation of a renormalizable model including only massive gauge
fields remains rather burning.

Recently in our papers (\cite{Sl1},\cite{Sl2}, \cite{Sl3}, \cite{QS}, \cite{QS1}, \cite{Sl4}\cite{Sl5})
 the new formulation of the Yang-Mills theory, which is free of the problem of Gribov ambiguity \cite{Gr} and therefore may be
used as in the framework of perturbation theory and beyond it, was proposed. It was shown that this formulation allows to overcome the theorem about the absence of soliton solutions of the classical equations of motion in the Yang-Mills theory and constructed explicit classical solutions corresponding to solitons.

In this paper we shall show that a suitable modification of this formalism possibly allows
to describe in the framework of renormalizable field theory massive gauge fields without introducing any additional observable scalar fields.

The paper is organized as follows. In the second chapter the Lagrangian describing the new theory is presented, and the problems related to renormalizability of the theory are discussed. In the third chapter the unitarity of the proposed model in the physical subspace is proven.  In the Conclusion the problems related to the application of the theory for the description of electro-weak models are briefly discussed.

\section{Formulation of the model}

 We start with the Lagrangian, describing, as shown in the paper \cite{Sl1} the massless Yang-Mills field
\begin{eqnarray}
L=-\frac{1}{4}F_{\mu\nu}^aF_{\mu\nu}^a+(D_\mu\varphi_+)^*(D_\mu\varphi_-)+(D_\mu\varphi_-)^*(D_\mu\varphi_+)+\nonumber\\
+(D _\mu b)^*(D _\mu e)+(D _\mu e)^*(D _\mu b)
\label{1}
\end{eqnarray}
where $\varphi_\pm$ are Bose fields, and $b,e$ are anti commuting elements of Grassman algebra. For brevity we consider the gauge group $SU(2)$, and the fields
$\varphi_\pm$ and  $b,e$ are the complex doublets, realizing its fundamental representation
\begin{eqnarray}
\varphi_\pm(x) \rightarrow \varphi_\pm^\omega=\Gamma[\omega(x)]\varphi_\pm(x)\nonumber\\
  \Gamma [\omega(x)]=\frac{ig}{2}\omega^a\tau^a
\label{2}
\end{eqnarray}
Analogous formulas are valid for the fields  $b,e$.

In the sector in which the ghost fields are not present in the asymptotic states one can in the path integral for the scattering matrix to integrate over the ghost fields $\varphi_\pm, b, e$ and verify that remaining integral coincides with the corresponding integral in the standard Yang-Mills theory. This simple proof does not work, if the asymptotic states contain the ghost particle.
Nevertheless we shall show below that the theory still may be unitary in the space which contains only observable particles provided it has some additional symmetry, which is absent in the standard formulation.

Our theory includes except for the nonabelian gauge fields the complex scalar doublets $\varphi_\pm, b,e$ which may be parametrized in terms of (anti)Hermitean components
\begin{equation}
\Phi=(\frac{i \Phi_1+\Phi_2}{\sqrt{2}}, \frac{\Phi_4-i \Phi_3}{\sqrt{2}}); \quad \Phi^*=(\frac{-i\Phi_1+\Phi_2}{\sqrt{2}}, \frac{\Phi_4+i\Phi_3}{\sqrt{2}})
\label{3}
\end{equation}
The components of the field $b$ we consider as antiHermitean.

In distinction of previous papers we assume that both fields $\varphi_\pm$ have nontrivial asymptotics
\begin{equation}
\varphi_\pm \rightarrow \frac{\hat{\alpha} m}{g}; \quad \varphi_\pm^* \rightarrow \frac{\hat{\alpha}m}{g},
\label{4}
\end{equation}
Let us introduce the variables $\tilde{\varphi}_\pm$ having zero asymptotic at infinity with the help of the formulas
\begin{equation}
\varphi_\pm  =\tilde{\varphi}_\pm+ \frac{\hat{\alpha} m}{g}; \quad \varphi_\pm^*=\tilde{\varphi}_\pm^*+ \frac{\hat{\alpha}m}{g},
\label{5}
\end{equation}
The fields $b, e$ decrease at infinity sufficiently fast.

In terms of the fields $\tilde{\varphi}_\pm$ the Lagrangian (\ref{1}) will look as follows
\begin{eqnarray}
L=-\frac{1}{4}F_{\mu\nu}^aF_{\mu\nu}^a+(D_\mu\tilde{\varphi}_+)^*(D_\mu\tilde{\varphi}_-)+(D_\mu\tilde{\varphi}_-)^*(D_\mu\tilde{\varphi}_+)+\nonumber\\
+\frac{m^2}{2}A_\mu^2+\frac{m}{g}(D_\mu\hat{\alpha}_+)^*(D_\mu\tilde{\varphi}_-)+\frac{m}{g}(D_\mu\hat{\alpha}_-)^*(D_\mu\tilde{\varphi}_+)+\nonumber\\
+h.c.+(D_\mu e)^*(D_\mu b)+(D_\mu b)^*(D_\mu e)
\label{6}
\end{eqnarray}
 In terms of (anti)Hermitean components they look as follows
\begin{eqnarray}
\delta A_\mu^a=\partial_\mu\eta^a+g\epsilon^{abc}A_\mu^b\eta^c\nonumber\\
\delta\tilde{\varphi}_\pm^a=\frac{\alpha_\pm m}{\sqrt{2}}\eta^a+\frac{g}{2}\epsilon^{abc}\tilde{\varphi}_\pm^b\eta^c+\frac{g}{2}\tilde{\varphi}_\pm^0\eta^a\nonumber\\
\delta\tilde{\varphi}_\pm^0=- \frac{g}{2}\tilde{\varphi}_\pm^a\eta^a\nonumber\\
\delta b^a=\frac{g}{2}\epsilon^{abc}b^b\eta^c+ \frac{g}{2}b^0\eta^a\nonumber\\
\delta e^a=\frac{g}{2}\epsilon^{abc}e^b\eta^c+ \frac{g}{2}e^0\eta^a\nonumber\\
\delta b^0=-\frac{g}{2}b^a\eta^a\nonumber\\
\delta e^0=-\frac{g}{2}e^a\eta^a
\label{7}
\end{eqnarray}
The remaining asymptotic fields do not change. One sees that in modified theory we have three fields changing by arbitrary functions. That means that in modified theory instead of one gauge fields we have three.
To quantize the Lagrangian (\ref{6}) one should choose a gauge. In this paper we are mainly interested in perturbation theory, where the ambiguity of quantization is absent. For this reason we choose the Lorentz gauge
$\partial_\mu A_mu^a=0$ and introduce the corresponding ghost fields. In this gauge the quadratic term, which is present in the Lagrangian (\ref{6}),and which has a form $A_\mu^a(x)\tau^a_{\gamma 2}\partial_\mu\varphi_\gamma(x)$
does not contribute and may be eliminated by redefinition of the Lagrange multiplier $\lambda$.

The effective action now looks as follows:
\begin{equation}
A_{ef}=\int dx[L+\lambda^a\partial_\mu A_\mu^a+i\partial_\mu\bar{c}D_\mu c].
\label{7c}
\end{equation}
where the Lagrangian $L$ is defined by the equation (\ref{6}).
Obviously the action  (\ref{7c}) corresponds to renormalizable theory, as all the propagators decrease at infinity as $k^{-2}$ and the interaction includes only the trilinear terms with one derivative and the four linear terms without derivatives.
So we succeeded to construct the theory which contains all necessary particles and which is renormalizable. Now we show that the is theory is unitary in the physical space.

\section{Unitarity in the physical space.}

The scattering problem in quantum theory is formulated in the following way. When $t \rightarrow \pm\infty$ the fields are described by the free equations of motion and in the process of scattering the probability
of transitions from the one asymptotic state to another one is measured. Of course some reshuffling of physical asymptotic states is possible. The asymptotic states are determined by the boundary conditions imposed on the scattering matrix. The unitarity of the model  will be proven if we can establish the equivalence  of our model in the physical subspace to the standard theory,
where the scattering matrix is unitary. So for the study of unitarity problem one can consider the asymptotic theory.

The asymptotic effective action (\ref{7c}) is asymptotically invariant  with respect to the supersymmetry transformations
\begin{eqnarray}
\delta\tilde{\varphi}_-=-ib\epsilon \nonumber\\
\delta{e}=\tilde{\varphi}_+ +const \nonumber\\
\delta{b}=\delta\tilde{\varphi}_+=0
\label{9}
\end{eqnarray}
These transformations are the asymptotic limit of the transformations which leave the effective action (\ref{7c}) unchanged. The transformations are obtained by the change of variables (\ref{5}) in the asymptotic effective action (\ref{7c}) and obviously leave the effective Lagrangian invariant, but they change the asymptotic of the field $e$. As these transformation change the asymptotical  behaviour of the field $e^0$,  one
can think that these transformations change the space where the theory is well defined, and therefore are not allowed. In  this section we shall demonstrate that the symmetry of the asymptotic action provides unitarity of the scattering matrix in the physical subspace, and the model, which we consider, has the scattering matrix, unitary in the subspace coinciding with the space of observables in the massive Yang-Mills theory.
This model may be used for the description of the massive nonabelian gauge theory in the framework of renormalizable model.

The action (\ref{6}) is invariant with respect to the gauge transformations. Hence it is invariant with respect to the BRST-transformations, which may be obtained from the transformations (\ref{7}) by changing $\eta^a$ to $ic^a\epsilon$, where $\epsilon$ is a constant anti commuting parameter. The effective action (\ref{7c}) is BRST invariant if
for the fields $\lambda, \bar{c}, c$ the following transformation law is adopted
\begin{equation}
\delta\lambda^a=0;  \quad   \delta\bar{c}^a=\lambda^a\epsilon;  \quad  \delta c^a=\frac{g}{2}\epsilon^{abd}c^bc^d \epsilon
\label{10}
\end{equation}


The natural choice of the physical states is given by the states, satisfying the following conditions
\begin{equation}
Q_B|\psi>_{ph}=0;   \quad   Q_S|\psi>_{ph}=0
\label{11}
\end{equation}
Here $Q_B$ is the BRST charge, conserving according to Neuther theorem due to invariance of the effective action with respect to BRST transformations. The operator $Q_S$ represents the similar charge
conserving due to invariance of the effective action with respect to the supersymmetry transformations .  But the supersymmetry transformations which leave the effective action invariant are not well defined.
According to the discussion, given above, it is sufficient to consider instead of equations (\ref{11}),the corresponding equations for the asymptotic states, that is the states, which are free, but provide the physical values of masses and charges. We assume that the symmetry of the theory is not spoiled by any anomaly.
The asymptotic conditions are imposed on the physical fields, the ghost fields are annihilated by the BRST Neuther charge $Q^{as}_B$ and the super symmetric charge $Q^{as}_S$.
\begin{equation}
Q_B^{as}|\psi>_{ph}^{as}=0; \quad Q_S^{as}|\psi>_{ph}^{as}=0;  \quad  [Q_B^{as}, Q_S^{as}]_+=0
\label{12}
\end{equation}
Here however we have a problem. The point is that the Neuther theorem guarantees the existence of the conserved current $\partial_\mu J_\mu=0$, if the action is invariant with respect to some
transformation of the fields and their derivatives, but does not guarantees the existence of the corresponding conserved charge. The problem of the correct definition of the invariance of the theory was considered in
 \cite{ IZ}, but here we just note that in this model there is a current which is well defined and
 asymptotically conserved.

We can replace $Q_S^{as}$ by the conserved, well defined operator
\begin{equation}
\tilde{Q}_S^{as}=\int d^3x(\partial_0\tilde{\varphi}_+^{as,\alpha}b^{as, \alpha}-\tilde{\varphi}_+^{as,\alpha}\partial_0b^{as, \alpha}+h.c.) \quad \alpha=0,1,2,3.
\label{15}
\end{equation}
$\alpha$ numerates hermitean and antihermitean components of $\tilde{\varphi}$ and $b,e$. We remind the reader that according to our convention  the field $b$ is antihermitean. One can see that this operator indeed exists as it is quadratic in the fields.
Conservation of the operator $\tilde{Q}_S^{as}$ may be easily checked directly, as the fields $b^{as}, \tilde {\varphi}_+^{as}$, satisfy the equations
\begin{equation}
\partial^2b^{as}=\partial^2\tilde{\varphi}_+^{as}=0
\label{16}
\end{equation}
It follows  from the hypothesis of adiabatic switching the interaction, that
\begin{equation}
\lim_{t \rightarrow \infty}U^{-1}(t)HU(t)=H^{as}
\label{17}
\end{equation}
where the operator $U(t)$ is given by the equation
\begin{equation}
U(t)=\exp\{-iH^{as}t\}
\label{17a}
\end{equation}
Let us introduce the operator $\tilde{Q}_S(t)$ which asymptotically coincides with $\tilde{Q}_S^{as}$.
\begin{equation}
\lim_{t\rightarrow \infty}U^{-1}\tilde{Q}_S(t)U=\tilde{Q}_S^{as}
\label{18}
\end{equation}
The following equality is the consequence of the fact that $[\tilde{Q}_S^{as},H_{as}]_-=0$
\begin{equation}
\lim_{t \rightarrow \infty}{[\tilde{Q}_S,H]_-}=0
\label{19}
\end{equation}
 The operator ${\tilde{Q}_S}$  commutes with  $H$  at $t \rightarrow \infty$.
The operator $U(t)$ may be introduced not only in the limit $t \rightarrow \infty$ but for the finite $t$ as well, although we need it only for $|t| \rightarrow \infty$. However the operator $Q_S(t)$ in general is nonlocal.The unitarity of the scattering matrix in the subspace of vectors annihilated by $\tilde{Q}_S^{as}$ follows from the formal calculation (\cite{F},\cite{FS})\begin{eqnarray}
S\tilde{Q}_S^{as}=\exp\{iH_{as}t''\}\exp\{-iH(t''-t')\}\exp\{-iH_{as}t'\}\tilde{Q}_S^{as}=\nonumber\\
=\exp\{iH_{as}t''\}\exp\{-iH(t''-t')\}\tilde{Q}_S\exp\{-iH_{as}t'\}=\nonumber\\
=\exp\{iH_{as}t''\}\tilde{Q}_S\exp\{-iH(t''-t')\}\exp\{-iH_{as}t'\}=\nonumber\\
=\tilde{Q}^{as}_S\exp\{iH_{as}t''\}\exp\{-iH(t''-t')\}\exp\{-iH_{as}t'\}=\tilde{Q}_S^{as}S
\label{20}
\end{eqnarray}
 Of course this proof does not exclude the possibility of some rearranging of the physical asymptotic states.
In this equation the limit $t''\rightarrow \infty; \quad t'\rightarrow -\infty$ is assumed.
As we saw the sufficient condition of the scattering matrix unitarity is the eq.(\ref{19}). Now we shall show that the physical subspace of our model essentially coincides with the space of the massive Yang-Mills theory in the standard formulation. For that purpose we introduce the operator $\hat{K}$ defined by the equation
\begin{equation}
[Q_S^{as},\hat{K}]_+\sim \hat{N}
\label{21}
\end{equation}
where $\hat{N}$ is the number operator of ghosts.
As the operator $\hat{K}$ we may take
\begin{equation}
\hat{K}=\int d^3x(\partial_0\tilde{\varphi}_-^{as,\alpha}e^{as, \alpha}-\tilde{\varphi}_-^{as,\alpha}\partial_0e^{as, \alpha}+h.c.) \quad \alpha=0,1,2,3.
\label{22}
\end{equation}
 The operator $Q_S^{as}$ is obviously nilpotent.
The operator $\hat{K}$ allows the similar representation.
In the equations (\ref{22}) the symbols $\tilde{\varphi}_\pm, b, e$ denote the creation and annihilation operators of the excitations of the corresponding fields, where the creation and annihilation operators are introduced by the formulas:
\begin{eqnarray}
\varphi(\mathbf{x,t})_\pm=\frac{1}{{2\pi}^{3/2}}\int[a^*_\pm (\mathbf{k})e^{-i\mathbf{kx}}\exp\{ik_0t\}+a_\pm(\mathbf{k})e^{i\mathbf{kx}}\exp\{-ik_0t\}]\frac{dk}{\sqrt{2\omega}}\nonumber\\
\pi_\pm (\mathbf{x,t})=\frac{1}{{2\pi}^{3/2}}\int[a_\pm^*(\mathbf{k})e^{-i\mathbf{kx}}\exp\{ik_0t\}-a_\pm (\mathbf{k})e^{i\mathbf{kx}}\exp\{-ik_0t\}]idk \frac{\sqrt{\omega}}{\sqrt{2}}\nonumber\\
\omega=k_0=\sqrt{\mathbf{k}}^2
\label{24}
\end{eqnarray}
We see that
\begin{equation}
[Q_S^{as},\hat{K}]_+=[Q_S^{as}\hat{K}+\hat{K}Q_S^{as}]
\label{25}
\end{equation}
Noting that any physical vector is annihilated by the operator $Q_S^{as}$ we see that any physical vector has the structure
\begin{equation}
|\psi>_{ph}=|\psi>_{\tilde{ph}}+|N>
\label{26}
\end{equation}
where the vector $|\psi>_{\tilde{ph}}$ contains the physical excitations of the massive Yang-Mills theory and the fields, corresponding to $\lambda$ and the Faddeev-Popov ghosts $\bar{c}, c$,  and the vector $|N>$ denotes the zero norm vector, containing $N$ ghost particles.

We shall illustrate this fact by the simple example. Let us consider the process two vector particle ${\rightarrow}$ two vector particle in the leading order of perturbation theory.
Let us consider the terms quadratic in $A_\mu^a$.  Consider the terms which are changed when the propagator of the vector particles is changed.
The second order terms are contained in the expression
\begin{equation}
 \sim \int dx \epsilon^{abc}f_{\mu \nu}^a A_{\mu}^b A_{\nu}^c \int dy \epsilon^{def}f_{\lambda \rho}^d A^e_{\lambda}A^f_{\rho}
\label{26a}
\end{equation}
Contribution of such terms to the scattering matrix is
\begin{eqnarray}
\sim \int dx\epsilon_{abc} A^a_\mu A^b_\nu(\partial_\mu A^c(x)_\nu-\partial_\nu A^c(x)_\mu)\int dy\nonumber\\
\epsilon_{def}A^d_\rho A^e_\sigma(\partial_\rho A^f(y)_\sigma-\partial_\sigma A^f(y)_\rho)
\label{26b}
\end{eqnarray}
We want to transform it to the transversal gauge, in which the propagators have a form
\begin{equation}
\tilde{D(k)}=\frac{g^{\mu\nu}-k^\mu k^\nu k^{-2}}{k^2-m^2}
\label{26c}
\end{equation}
Some terms,  which contain propagators $F_{\mu \nu}$ ,will not change under this transformation. We remind that for the free field $\partial_\mu A^a_
\mu=0$ automatically. The difference between the original expression and the same expression in the transversal gauge is equal to
\begin{eqnarray}
\epsilon^{abc}\partial _\mu A_\nu^b(x)\partial_\nu A_\mu^c(x)\epsilon^{ade}\partial_\rho
A^d_\sigma(y)\partial_\sigma A_\rho^e(y)D_c^0(x-y)=\nonumber\\
(\partial_\mu A_\nu^b(x)\partial_\rho A_\sigma^b(y))(\partial_\nu A_\mu^c(x)D_c^0(x-y)-(\sigma \rightarrow \rho, \rho \rightarrow \sigma)=0
\label{26d}
\end{eqnarray}
2. Now we shall show that additional terms, which are present in the Lagrangian (\ref{6}), generates the term $\tilde{S_1}^*(x)
\tilde{S_1}(y)+\tilde{S_1}^*(y)\tilde{S_1}(x)$, which exactly compensates the term $\tilde{S_2}^*+\tilde{S_2}$ and unitarity condition acquires the standard form. ($\tilde{S_i}$ describes the process which is due to additional terms in the Lagrangian (\ref{6}))
\begin{equation}
S_2^*(x,y)+S_2(x,y)+S_1^*(x)S_1(y)+S_1^*(y)S_1(x)=0.
\label{26m}
\end{equation}
 In fact the disappearance of the pole at zero mass was demonstrated here.

Now we shall show that the fields $\lambda$ and Faddeev-Popov ghosts also decouple. It is necessary however to have in mind that the proof given above refers only to asymptotic states.
We  proved that the fields $\varphi_\pm,b, e$ enter only via zero norm vectors and therefore do not contribute to the observable matrix elements. However for explicit renormalisability it may be necessary to preserve some ghost fields. For example without breaking the renormalisability we cannot omit the fields $\lambda$ in the intermediate states, as only the special combination of fields $\tilde{A}_\mu, \lambda$ has sufficiently fast decreasing propagator.

The effective action (\ref{7c}) is invariant with respect to the transformation, generated by the generator of BRST transformations $Q_B$.
In the Lagrangian, describing the fields $\lambda, \varphi_\pm, b, e, A_\mu, c, \bar{c}$ we shall make the change of variables
\begin{equation}
A^a_\mu \rightarrow\tilde{A}^a_\mu-\frac{1}{m}\partial_\mu \lambda^a
\label{30}
\end{equation}
This change eliminates the cross term for the fields $\lambda^a, A_\mu^a$, and the effective Lagrangian takes the form
\begin{eqnarray}a
L_{ef}= -\frac{1}{4}\tilde{F_{\mu\nu}}^a\tilde{F_{\mu\nu}}^a+\frac{m^2}{2}(\tilde{A}_\mu^a)^2+(\partial_\mu \lambda^a)^2+\partial_\mu {\varphi}^*_+\partial_\mu{\varphi_-}+\partial_\mu(\bar{c})\partial_\mu(c)\nonumber\\
+\frac{m}{g}(D_\mu \tilde \alpha_+)^*(\partial_\mu\tilde{\varphi})_-+\frac{m}{g}(D_\mu\tilde{\alpha})_-^*(\partial_\mu\tilde{\varphi}_+)+\partial_\mu {b}^*\partial_\mu e+ h.c.+L_{int}(\tilde{A}_\mu^a-\frac{1}{m} \partial_\mu \lambda^a)
\label{31}
\end{eqnarray}
Here $\tilde{F_{\mu\nu}}$ denotes the free part of the curvature tensor. We wrote explicitly the asymptotic effective Lagrangian, and denote all other terms as $L_{int}$. As the field $\lambda$ does not change under the BRST-transformation, the transformations (\ref{7c}, \ref{10}) preserve the symmetry of the asymptotic action .
Note that the field $\lambda$ enters with the factor $\frac{1}{2}$. Using the equation of motion for the fields $\lambda, c, \tilde{A}_\mu,b, e$ it is easy to show that the operator $Q_B^{as}$ is conserved.
Therefore it commutes with $H^{as}$.
\begin{equation}
[Q_B^{as},H^{as}]=0
\label{32}
\end{equation}
As we explained above, the equation (\ref{32}) leads to equality
\begin{equation}
[Q_B^{as},H^{as}]_{|t|\rightarrow \infty} \rightarrow[Q_B,H]_-
\label{32a}
\end{equation}
where
\begin{equation}
(Q_B)_{|t| \rightarrow \infty}=U^{-1}Q_B^{as}U
\label{33}
\end{equation}
which guarantees the unitarity of the scattering matrix in the subspace of states annihilated by $Q_B^{as}$.

The essential coincidence of this  subspace with the space of the physical states in the standard formulation of the massiveYang-Mills theory may be established in a way similar to the one used for the states annihilated by $Q_S^{as}$.
According to the discussion above the operator $Q_B^{as}$ may be written in the form
\begin{equation}
Q_B^{as}= \int d^3k [a_{\lambda^a}(\mathbf{k})^*a_{c^a}(\mathbf{k}) -a_{\lambda^a}(\mathbf{k})a_{c^a}^*(\mathbf{k})]
\label{34}
\end{equation}
The conditions about notations are the same as in the corresponding equation for $Q_S^{as}$.

Let us introduce the operator
\begin{equation}
\hat{M}=\int d^3k[a_{\lambda^a}(\mathbf{k})^*a_{\bar{c}^a}(\mathbf{k})-a_{\lambda^a}(\mathbf{k})a_{\bar{c}^a}(\mathbf{k})^*]
\label{35}
\end{equation}
where the operators $a_{\lambda^a}(\mathbf{k})^*, a_{\lambda^a}(\mathbf{k}), a_{c}^a(\mathbf{k}), a_{\bar{c}}^a(\mathbf{k})$ satisfy the usual commutation relations.

It is easy to see that
\begin{equation}
[Q_B^{as},\hat{M}]_+\sim \int d^3k[2a_{\lambda^a}^*(\mathbf{k})a_{\lambda^a}(\mathbf{k})-a_{\bar{c}^a}^*(\mathbf{k})a_{c^a}(\mathbf{k})-a_{c^a}^*(\mathbf{k})a_{\bar{c}^a}(\mathbf{k})]
\label{36}
\end{equation}
The square of this operator is zero. Here we took into account, that
the number of particles corresponding to $c^a$ is equal to the number of particles corresponding to $\bar{c}^a$.
Hence for any $N$
\begin{equation}
Q_B^{as}\hat{M}+\hat{M}Q_B^{as} \sim \hat{N}
\label{37}
\end{equation}
Acting on any physical vector, annihilated by the operator $Q_B^{as}$, the second term in the eq.(\ref{37}) gives zero and any vector, annihilated by the first term has zero norm.
Therefore any vector annihilated by both operators $Q_B, Q_S$ has the structure
\begin{equation}
|\psi>_{ph}=|\psi>_{YM}+|N>
\label{38}
\end{equation}
where $ |\psi>_{YM}$ is the physical vector in the standard formulation of a massive nonabelian gauge theory and $|N>$ is a zero norm vector.
To have a complete equivalence of our model to the standard massive gauge theory we must factorize the space of vectors, annihilated by the operators $Q_B, Q_S$  with respect to zero norm vectors.

\section{Conclusion}

In this paper we presented the model which does not require for renormalizability the presence of observable scalar fields. Nevertheless these additional fields may be introduced into the model without any problems.
One can ask how this paper agrees with the paper \cite{CLT}, where it is claimed that the only massive renormalisable model is the Higgs model. Our paper does not contradict this work. It describes the different renormalizable theory. In particular there is no gauge where the free propagators are given by the equation (2) of the reference \cite{CLT}.
There is no gauge, in which we have only Yang-Mills particles. The Lagrangian (\ref{6}) differs from the standard Yang-Mills Lagrangian.

\section{Aknowledgements}.
 This work was done at Steklov Mathematical Institute and supported by the Russian Scientific Foundation.
I wish to express my deep gratitude to D.Bykov, L.Faddeev,  for useful discussions. My special thanks to A.Quadri, whose criticism helped me a lot.

\end{document}